\documentstyle[12pt,epsfig]{article}
\def \medio  {\baselineskip= 1.5 \normalbaselineskip}

\def \as {\alpha_s}

\newcommand{\titul}[1] {\begin{center}{\large\bf #1 } \end{center}\vskip 1.cm}
\newcommand{\autor}[1] {\begin {center} {\large \lineskip .5em #1 }
                        \end   {center} }
\newcommand{\lugar}[1] {\begin{center} {\it #1} \end{center}}
\newcommand{\abstr}[1] {{\begin{center} \vskip .5cm {\bf Abstract
                        \vspace{0pt}} \end{center}}\begin{quote} #1
                        \end{quote}}

\newcommand{\aas}{\alpha_s}

\newcommand{\bea}{\begin{eqnarray}}
\newcommand{\eea}{\end{eqnarray}}
\begin{document}
\begin{titlepage}

\begin{flushright} {
} \end{flushright}

\titul{ The 
longitudinal structure function $F_L$:\\ perturbative QCD 
and $k_T$-factorization \\ versus experimental data at fixed $W$
}

\autor{
A.V. Kotikov
\footnote{
on leave of absence from the Joint Institute for Nuclear Research,
 141980 Dubna, Moscow region, Russia}
}
\lugar{
  Institut f\"ur Theoretische  Teilchenphysik\\
  Universit\"at Karlsruhe\\
  D-76128 Karlsruhe, Germany
}
\autor{A.V. Lipatov}
\lugar{Department of Physics \\
Lomonosov Moscow State University\\
119899 Moscow, Russia}
\autor{N.P. Zotov}
\lugar{
Skobeltsyn Institute of Nuclear Physics \\
Lomonosov Moscow State University\\
119992 Moscow, Russia}
\abstr{
\medio
We use results for the structure function $F_L$ 
for a gluon target having nonzero transverse
momentum square at order $\alpha _s$, obtained in our previous paper, 
to compare with recent H1 experimental data for $F_L$ at fixed $W$ values
and with collinear GRV predictions at LO and NLO approximation.

\vskip 0.5cm

PACS number(s): 13.60.Hb, 12.38.Bx, 13.15.Dk

}
\end{titlepage}
\newpage

\pagestyle{plain}

The longitudinal structure function (SF) $F_L(x,Q^2)$ 
is a very sensitive QCD characteristic and is directly connected to the
gluon content of the proton.   
It is equal to zero in the parton model with spin$-1/2$ partons
and has got nonzero values in the framework of perturbative Quantum 
Chromodynamics.
The perturbative QCD, however, leads to a quite controversal results.
At the leading order (LO) approximation $F_L$ amounts to about $10\div 20 
\%$ of the corresponding $F_2$ values at large
$Q^2$ range and, thus, it has got quite 
large contributions at low $x$ range.
The next-to-leading order (NLO) corrections to the longitudinal 
coefficient function are large and negative at small $x$ 
\cite{Neerven}-\cite{Rsmallx}
and can lead to negative $F_L$ values at 
low $x$ and low $Q^2$ values (see \cite{Rsmallx,Thorne02}).
Negative $F_L$ values demonstrate a limitations of the applicability of
perturbation theory and the necessity of a resummation procedure, 
that leads to 
coupling constant scale higher than $Q^2$ 
(see \cite{Rsmallx}, \cite{DoShi}-\cite{Salam}).
 
The experimental extraction of $F_L$ data requires a
rather cumbersome procedure, especially at small values of $x$
(see \cite{1.5}, for example).
Recently, however,  there have been presented 
new precise preliminary H1 data  \cite{H1rev} on the longitudinal SF
$F_L$, 
which have probed the small-$x$ region $ 10^{-5} \leq x \leq 10^{-2}$.

In this paper 
 the standard perturbative QCD formulas and also the so called
$k_T$-factorization approach \cite{CaCiHa}
based on Balitsky-Fadin-Kuraev-Lipatov (BFKL) dynamics \cite{BFKL}
(see also recent review \cite{Andersson} and references therein)
is used for the analysis of the above data.
The perturbative QCD approach is called  hereafter as collinear
approximation and applied at LO and NLO levels using GRV
parameterizations for partion densities (see \cite{GRV}). The corresponding
coefficient functions are taken from the papers \cite{Guillen,Neerven}.

In the framework of the $k_T$-factorization approach, which 
 is of primary consideration in our paper, a study of the 
longitudinal SF $F_L$ has been done firstly in Ref. \cite{CaHa},
where the small $x$ asymptotics of $F_L$ has been obtained analytically
using the BFKL results for  the Mellin transform of the
unintegrated gluon distribution and the longitudinal Wilson coefficient 
functions for the full perturbative series has 
been calculated  at asymptotically small $x$ values. 
In this note we follow a more phenomenological approach \cite{KLZ} 
where we analyzed $F_L$ data in a broader range at small $x$ and, thus,
we use parameterizations
of the unintegrated gluon distribution function $\Phi_g(x,k^2_{\bot})$
(see Ref. \cite{Andersson}).

A similar study has been already done 
\footnote{Note that the studies of the $F_L$ structure function
in the framework of the $k_T$-factorization have been done also in
\cite{Blumlein93,BaKwSt}.}
in our paper \cite{KLZ} using previous H1 data \cite{H1FL97}.
The recent H1 preliminary experimental data \cite{H1rev} is essentially
more precise, that stimulates the present additional study.\\

{\bf 1.}~
The  unintegrated gluon distribution $\Phi_g(x,k^2_{\bot})$ 
($f_g$ is the (integrated) gluon distribution in the proton multiplied
by $x$ and $k_{\bot}$ is the transverse part of the gluon 4-momentum 
$k^{\mu}$)
 \begin{eqnarray}
f_{g}(x,Q^2) ~=~ \int^{Q^2}dk^2_{\bot}
\, \Phi_g(x,k^2_{\bot}) 
~~~~~\mbox{(hereafter } 
~k^2=-k^2_{\bot} \mbox{)}
\label{1}
 \end{eqnarray}
is the basic dynamical quantity in 
the $k_T$-factorization approach \footnote{
In our previous analysis \cite{KLPZ} we have shown
that the property
$k^2=-k^2_{\bot}$ leads to
the equality of the Bjorken $x$ value  
in the standard renormalization-group approach 
and in the Sudakov one.}.
It satisfies the BFKL equation \cite{BFKL}. 

Then, in the $k_T$-factorization
the SF $F_{2,L}(x,Q^2)$ are driven at small $x$ primarily
by gluons and are related in the following way to the unintegrated 
distribution $\Phi_g(x,k^2_{\bot})$: 
\begin{eqnarray}
F_{2,L}(x,Q^2) ~=~\int^1_{x} \frac{dz}{z} \int^{Q^2} 
dk^2_{\bot} \sum_{i=u,d,s,c} e^2_i
\cdot \hat C^g_{2,L}(x/z,Q^2,m_i^2,k^2_{\bot})~ \Phi_g(z, k^2_{\bot}), 
 \label{d1}
\end{eqnarray}
where $e^2_i$ are charge squares of active quarks.

The functions $\hat C^g_{2,L}(x,Q^2,m_i^2,k^2_{\bot})$ 
can be regarded as  SF of the 
off-shell gluons with virtuality $k^2_{\bot}$ (hereafter we call them
{\it hard structure functions } by analogy with similar 
relations between cross-sections and hard 
cross-sections). They are described by the sum of the quark 
box (and crossed box) diagram contribution to the 
photon-gluon interaction 
(see, for example, Fig. 1 in \cite{KLPZ} and \cite{KLZ}). \\

{\bf 2.}~
Notice that the $k^2_{\bot}$-integral in Eqs. (\ref{1}) and (\ref{d1})
can be divergent at lower limit,
at least for some parameterizations of $\Phi_g(x,k^2_{\bot})$.
To overcome the problem we change the low $Q^2$ asymptotics of
the QCD coupling constant within hard structure functions.
We apply here two models: the ``freezing'' procedure and
Shirkov-Solovtsov analytization.

The ``freezing'' of the strong coupling constant is very popular 
phenomenological model for infrared behavior of 
$\as(Q^2)$.
The ``freezing'' can be done in the hard way and in the soft way. 

In the hard case (see \cite{NikoZa}, for example), the strong coupling 
constant itself is modified: it is
taken to be constant at all $Q^2$ values less then some $Q^2_0$, i.e.
$\aas(Q^2)~=~ \aas(Q^2_0)$, if $Q^2 \leq Q^2_0$.

In the soft case (see \cite{BaKwSt}, for example), the subject of 
the modification is the argument of the strong coupling constant. It
contains the shift $Q^2 \to Q^2 + M^2$, where $M$ is an additional
scale, which strongly modifies the infrared $\alpha_s$ properties.
For massless produced quarks, $\rho$-meson mass $m_{\rho}$ is usually 
taken  
as the $M$ value, i.e. $M=m_{\rho}$.
In the case of massive quarks with  mass $m_i$, the $M=2m_{i}$ value
is usually used.
Below we will use the soft version of ``freezing'' procedure.

Shirkov and Solovtsov proposed \cite{ShiSo} a 
procedure of analytization of the strong coupling constant $\as(Q^2)$, 
which leads to a new strong analytical coupling constant $a_{an}(Q^2)$ 
having 
nonstandard infrared properties. We are not in position to discuss here
theoretical 
aspects of the procedure and use only the final formulae for the 
analytical coupling constant $a_{an}(Q^2)$. They have the following form

\begin{eqnarray}
\frac{a_{an}(Q^2)}{4\pi} ~=~ \frac{1}{\beta_0} 
\left[
\frac{1}{\ln(Q^2/\Lambda^2)} + \frac{\Lambda^2}{\Lambda^2-Q^2}
\right]
\label{cc:LO}
 \end{eqnarray}
in the LO approximation and

\begin{eqnarray}
\frac{a_{an}(Q^2)}{4\pi} ~=~ \frac{1}{\beta_0} 
\left[
\frac{1}{\ln(Q^2/\Lambda^2) + b_1 \ln[1+ \ln(Q^2/\Lambda^2)/b_1]} 
+ \frac{1}{2} \, \frac{\Lambda^2}{\Lambda^2-Q^2} - 
\frac{\Lambda^2}{Q^2} \, C_1
\right],
\label{cc:NLO1}
\end{eqnarray}
in the NLO approximation, where
$\beta_0$ and $\beta_1$ are the two first terms in the
$\alpha_s$-expansion of $\beta$-function and $b_1=\beta_1/\beta^2_0$.
The constant $C_1=0.0354$ is very small.

The first terms in the r.h.s. of Eqs. (\ref{cc:LO}) and (\ref{cc:NLO1})
are the standard LO and NLO representations for $\alpha_s(Q^2)$. The
additional  terms modify its infrared properties.

Note that numerically both infrared transformations, the ``freezing'' 
procedure and Shirkov-Solovtsov analytization, lead to very close results
(see below Fig. 1 and also Ref. \cite{IllaKo} and discussion therein).

\begin{figure}[htb]
\begin{center}
\epsfig{figure= 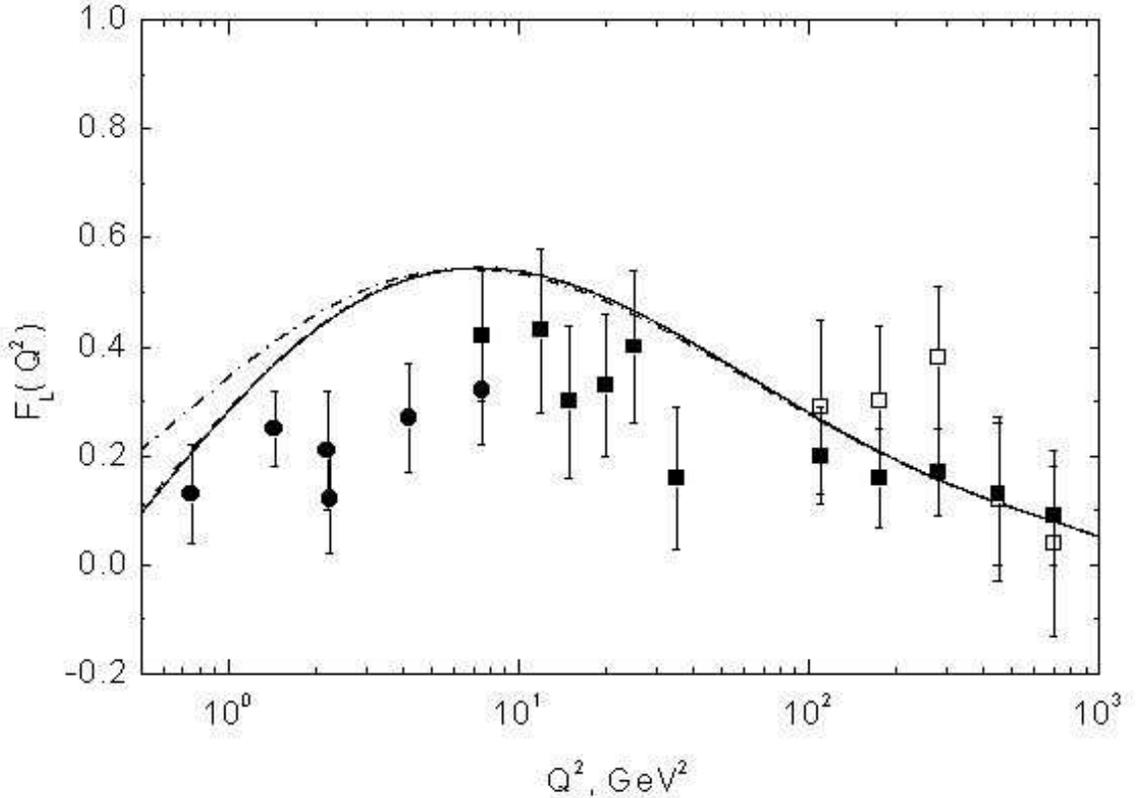,width=17cm,height=13cm}
\end{center}
\caption{$Q^2$ dependence of $F_L(x, Q^2)$ (at fixed $W$ = 276 GeV).
The H1 preliminary, $e^+p$ and $e^-p$ experimental data are shown as 
the black points, black and white squares, respectively (see \cite{H1rev}).
Theoretical curves obtained in the $k_T-$factorization approach
with the JB unintegrated gluon distribution: solid curve corresponds to
"frozen" coupling constant, dashed curve - analytical
 coupling constant, dash-dotted - "frozen" argument of the 
unintegrated gluon distribution function.}
\label{fig1}
\end{figure}
 
{\bf 3.}~ As it was already noted above,
the purpose of the paper is to describe  new preliminary 
H1 experimental data 
for the longitudinal SF $F_{L}(x,Q^2)$ using our
calculations of the hard SF
$\hat C^g_{2,L}(x,Q^2,m^2,k^2_{\bot})$ given in our previous study
\cite{KLPZ} and infrared modifications of
$\aas(Q^2)$, explained above.
For the unintegrated gluon distribution $\Phi (x, k^2_{\bot}, Q_0^2)$
we use the so called Blumlein's parametrization (JB)
\cite{Blumlein}. 
Note that there are also several other popular parameterizations, which
give quite similar results  excepting, 
perhaps, the contributions from the small $k_{\bot}^2$-range: 
$k_{\bot}^2 \leq 1$ GeV$^2$ 
(see Ref. \cite{Andersson} and references therein).

The JB form depends strongly on the Pomeron intercept value.
In different models the Pomeron intercept 
has different values (see \cite{Kaidalov}). So, in
our calculations we apply the H1 parameterization \cite{H1slope} based
on the corresponding H1 data, which
are in good agreement with
perturbative QCD (see Refs. \cite{H1slope,KoPa03}).

We calculate the SF $F_L$ as the sum of two types of contributions:
the charm quark one $F^c_L$ and the light quark one $F^{l}_L$:
\bea 
F_L  ~=~ F^{l}_L + F^{c}_L
\label{nu1}
\eea

For the $F^{l}_L$ part we use the massless limit of hard SF (see 
\cite{KLPZ,KLZ}).
We always use $f=4$ in  our fits, because our results depend very
weakly on the exact $f$ value (for similar results see 
fits of experimental data in \cite{KriKo}
and discussions therein). The weak dependence comes from two basic
properties. Firstly, the charm part of $F_L$, $F_L^c$, is quite small at
the considered $Q^2$ values (see Ref. \cite{KLZ} for the $F_L^c$ study).
Secondly, the strong coupling constant very weakly depends  on $f$ 
because of the corresponding relations between $\Lambda$ values at 
different $f$ (see \cite{Chetyrkin}).

\begin{figure}[htb]
\begin{center}
\epsfig{figure= 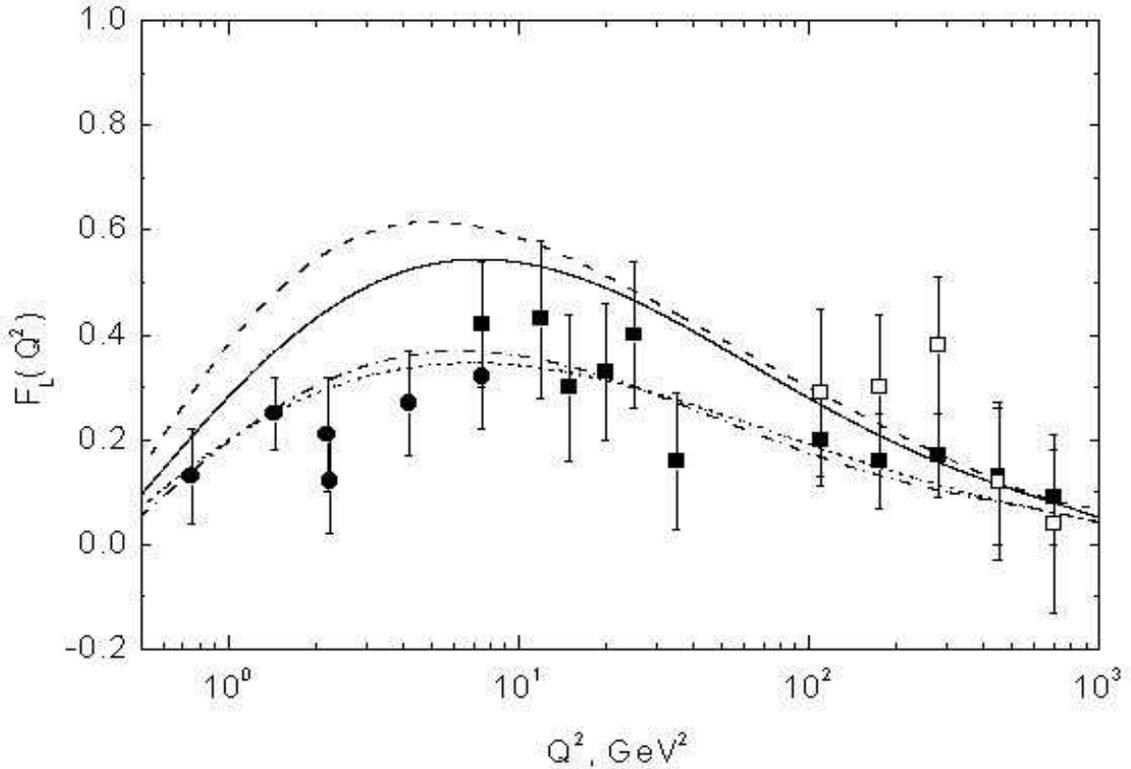,width=17cm,height=12.5cm}
\end{center}
\caption{$Q^2$ dependence of $F_L(x, Q^2)$ (at fixed $W$ = 276 GeV).
The experimental points are as in Fig. 1.
Solid curve is the result of the $k_T-$factorization approach
with the JB unintegrated gluon distribution and
"frozen" coupling constant, dashed curve - the GRV LO calculations,
dash-dotted curve - the GRV NLO  calculations, dotted curve
- the result of the GRV LO calculations    
 with  
$\mu^2 = 127Q^2$.}
\label{fig2}
\end{figure}

In Fig. 1 we show  the SF $F_L$ with ``frozen'' and analytical
coupling constants, respectively, as a function of $Q^2$ for fixed
$W$  in comparison with H1 experimental data sets (see \cite{H1rev}).
The results are mostly coincide with each other. They are presented as 
bold and
dashed curves, which cannot be really resolved in the figure.

 The dash-dotted curve shows the results obtained with ``frozen''
argument of the unintegrated gluon density. 
The difference between the bold and dash-dotted
lines is not so big, that demonstrates the unimportance of the infrared
modifications of the density argument. Below we only restrict ourselves 
only to the modification of the  
argument in the strong coupling constant entering the hard structure
function.

 Fig. 2 contains the same bold curve as Fig. 1 and shows also the
collinear results for $F_L$ values. We use the popular GRV
parameterizations \cite{GRV} at LO and NLO approximations. 
The $k_T$-factorization results lie between the collinear ones, that  
demonstrates clearly the particular resummation of high-order collinear
contributions at small $x$ values in the $k_T$-factorization approach.

We also see  exellent agreement between the experimental data and
collinear approach with GRV parton densities at NLO approximation. The
NLO corrections are large and negative and decrease the $F_L$ value 
by an approximate factor of 2 at $Q^2 < 10$ GeV$^2$.

In Figs. 1 and 2, our $k_T$-factorization
 results are in good agreement with the data for large and small
parts of the $Q^2$ range. We have, however, some disagreement
 between the data 
and theoretical predictions at $Q^2 \sim 3$ GeV$^2$. The disagreement 
exists in both cases: for collinear QCD approach at the LO 
approximation
and for $k_T$-factorization.

Comparing these results with Fig. 4 of Lobodzinska's talk in Ref. 
\cite{H1rev} we conclude 
that the disagreement comes from the usage of the LO approximation.
Unfortunately, at the moment in the $k_T$-factorization approach only 
the LO terms are available. The calculation of the NLO corrections is  
a very complicated problem (see \cite{Jung} and discussion therein).

A rough estimation of the NLO corrections in the $k_T$-factorization
approach can be done in the following way.
Consider first the BFKL approach. A popular resummation of the
NLO corrections is done in \cite{BFKLP} at some approximation. Ref.
\cite{BFKLP} demonstrates, that is 
the basic effect of the NLO  corrections, that is
the strong rise of the $\alpha_s$ argument from $Q^2$ to $Q^2_{eff} =
K \cdot Q^2$, where $K=127$, i.e. $K>>1$, which is in agreement with
\cite{Rsmallx}, \cite{DoShi} and \cite{Salam}.

\begin{figure}[htb]
\begin{center}
\epsfig{figure= 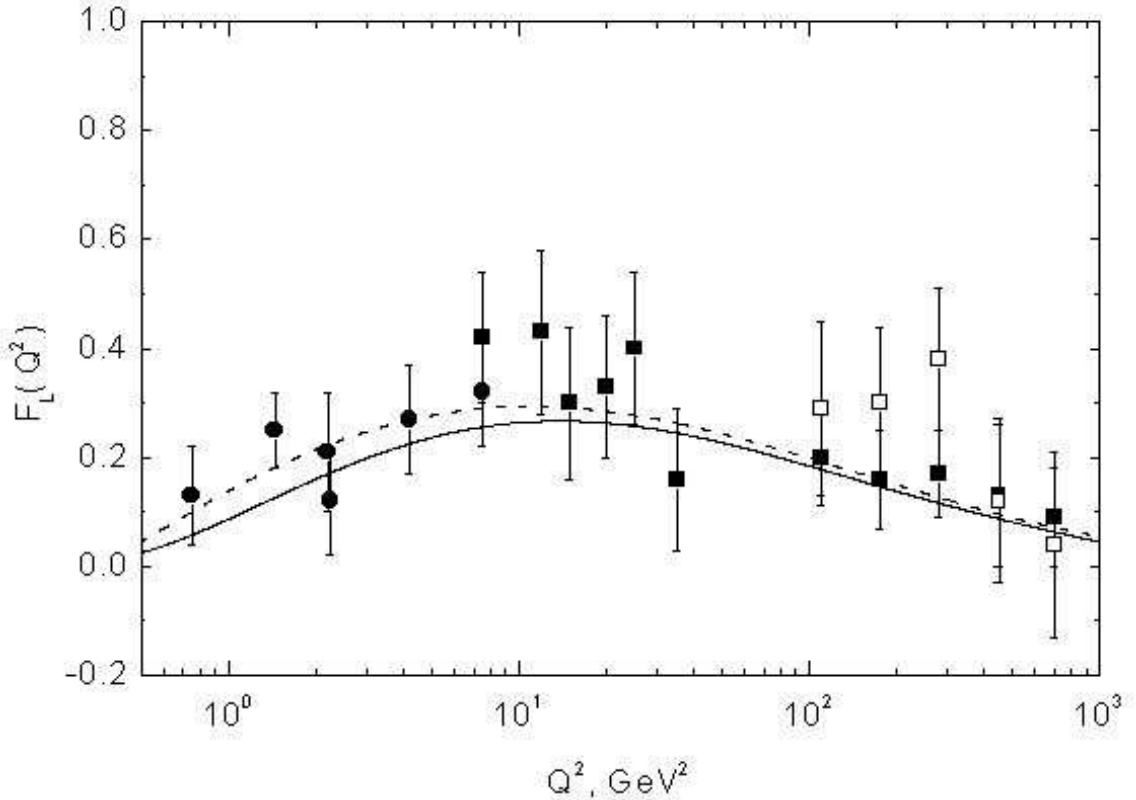,width=17cm,height=13cm}
\end{center}
\caption{$Q^2$ dependence of $F_L(x, Q^2)$ (at fixed $W$ = 276 GeV).
The experimental points are as in Fig. 1.
Solid curve is the result of the $k_T-$factorization approach
with the JB unintegrated gluon distribution and
$\mu^2 = M_Z^2$, dashed curve - the GRV LO 
calculations at $\mu^2 = M_Z^2$.}
\label{fig3}
\end{figure}

The use of the effective argument $Q^2_{eff}$ in the DGLAP approach at LO
approximation leads to results which are very close to the ones 
obtained in the case of
NLO approximation: see the dot-dashed and dotted curves in Fig. 2. 
Thus, we hope that the effective
argument represents the basic effect of the NLO  corrections in the
framework of the $k_T$-factorization, which in some
 sense lies between the DGLAP
and BFKL approaches as it was noted above already.

The necessity of large effective arguments is also demonstrated in
Fig. 3, where we show the $k_T$-factorization and collinear results for
nonrunning coupling constant. Its argument is fixed at $Q^2=M_Z^2$ giving
$\alpha_s \approx 0.118$ (see \cite{Bethke}), i.e. the considered argument is
larger than the most part of the $Q^2$-values of the considered 
experimental data. 
\footnote{The study is also initiated by
conversation with L.L\"onnblad, we thank him.}

The results obtained in the $k_T$-factorization and collinear approaches
based on $Q^2_{eff}$ argument are presented in Fig. 4. In comparison
with the ones shown in Fig. 1, they are close to each other because the 
effective argument
is essentially larger than the $Q^2$ value. There is very good agreement
between the experimental data and both theoretical approaches.

Moreover, we also present in Fig.4 the $F_L$ results based on the
$R_{world}$-parameterization for the $R=\sigma_L/\sigma_T$ ratio (see
\cite{SLAC}) (because $F_L=F_2 R/(1+R)$), 
improved in \cite{CCFR,CCFRr} for low $Q^2$ values and the 
parameterization of $F_2$ data used in the our previous paper
\cite{KLZ}. The results are in good agreement with other
theoretical predictions  as well as with experimental data.\\

\begin{figure}[htb]
\begin{center}
\epsfig{figure= 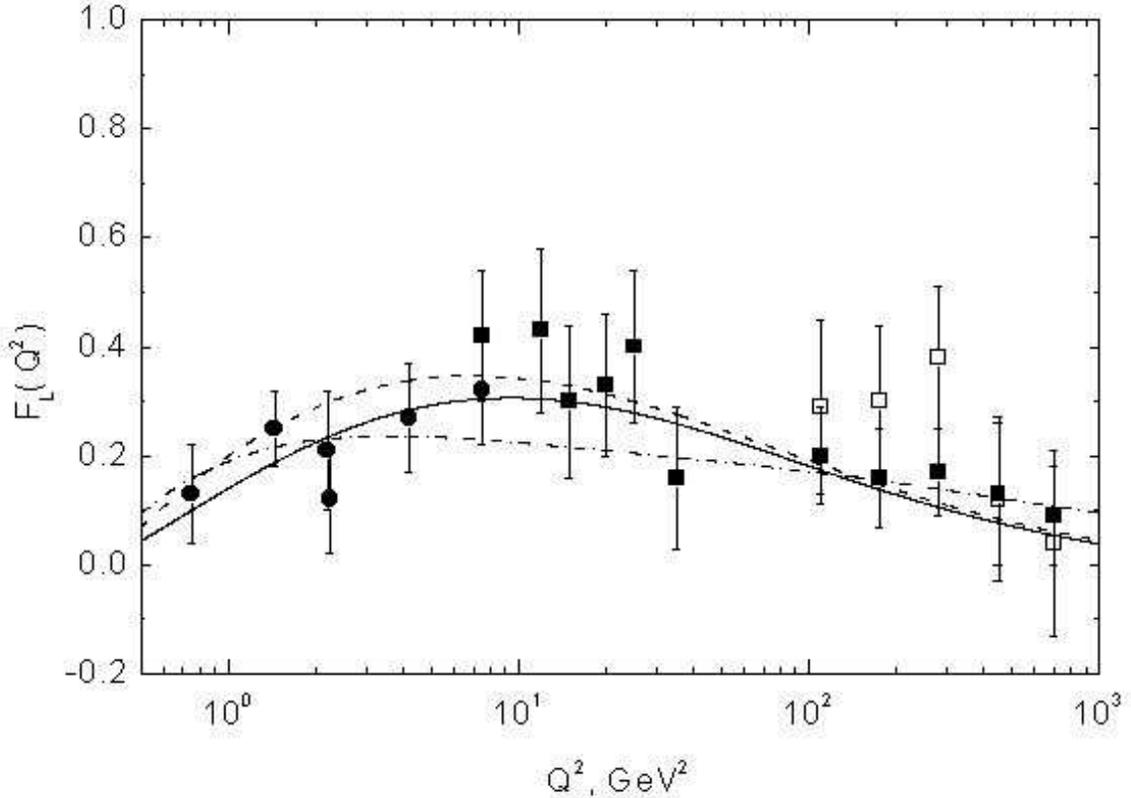,width=17cm,height=13cm}
\end{center}
\caption{$Q^2$ dependence of $F_L(x, Q^2)$ (at fixed $W$ = 276 GeV).
The experimental points are as in Fig. 1.
Solid curve is the result of the $k_T-$factorization approach  
with the JB unintegrated gluon distribution and
at $\mu^2 = 127Q^2$, dashed curve - the GRV LO
calculations at $\mu^2 = 127Q^2$, dash-dotted curve - from the 
$R_{world}$-parametrization.}
\label{fig4}
\end{figure}

{\bf 4.}~~{\it Resume}.
 In the framework of $k_T$-factorization we have applied 
the results of the calculation of the perturbative
parts for the structure functions $F_L$ and $F_L^c$ 
for a gluon target, 
having nonzero momentum square, in the process 
of photon-gluon fusion \cite{KLPZ, KLZ}
to the analysis of recent H1 preliminary data.
The perturbative QCD predictions are presented also at LO and NLO 
approximations.

We have found very good agreement between the experimental data and
collinear results based on GRV parameterization at NLO approximation.
The LO collinear and $k_T$-factorization results show disagreement
with the data at some $Q^2$ values. We argued that the disagreement
comes from the absence of the NLO corrections in the framework of 
the $k_T$-factorization.
We modeled these NLO corrections by
choosing  large effective argument of the strong coupling constant and
argued for our choice.
The effective corrections significantly improve the agreement with the H1 
data under consideration.\\

\hspace{1cm}  
{\bf Acknowledgements}    \vspace{0.5cm}

\normalsize{}

We thank S.P. Baranov for careful reading of manuscript and useful 
remarks.
The our study is supported in part by the RFBR grant.
One of the authors (A.V.K.) is supported in part by 
Alexander von Humboldt fellowship. 
A.V.L. is supported in part by INTAS YSF-2002 grant $N^o$ 399 and
"Dinastiya" Fundation.
N.P.Z. also acknowledge L. J\"onsson for discussion of the H1 data 
\cite{H1rev} and the support of Crafoord Fundation (Sweden).


\end{document}